\documentclass[10pt, twoside]{article}
\usepackage{epsfig}
\usepackage[lflt]{floatflt}

\begin{document}

\begin{center}
{\Large Motion of a falling object}
\vskip 0.5cm

{\tt Kalimullah, Madhur Garg, P. Arun \\
Department of Physics \& Electronics, S.G.T.B. Khalsa College,\\
University of Delhi, Delhi - 110 007, India\\
\vskip 0.2cm
\&\\
\vskip 0.2cm
F.M.S. Lima\\
Instituto de Fisica\\
Universidade de Brasilia\\
P.O. Box 04455\\
70919-970 Brasilia-DF\\
Brazil.\\
}

\vskip 0.5cm
{\bf Abstract}
\end{center}

\par {\it A simple setup was assembled to study the motion of an object
while it falls. The setup was used to determine the instantaneous velocity, 
terminal velocity and acceleration due to gravity. Also, since the whole 
project was done within \$20 it can easily be popularized.}\\
\vskip 0.25cm

\section{Introduction}

\par One of the first topic students of physics learn is the concept of
bodies falling towards the center of the earth with an acceleration (g). This 
acceleration due to gravity is usually determined in undergraduate
laboratories by measuring the oscillation time period of a simple pendulum.
Some students do wonder as to why acceleration due to gravity is not just 
determined by measuring the time taken by a freely falling body to fall through 
a given height. Trivial calculation using the equation 
\begin{eqnarray}
s=ut+{1 \over 2}gt^2\label{eq1}
\end{eqnarray}
shows that the body has to drop across an appreciable height to increase the
body's time of flight. This is required to minimize the error in
measurement of the fall's time. Historically this would explain why Galileo
did his experiment from the Tower of Pisa. Over the last three decades
various contraptions have been designed to measure the time of flight in
laboratory conditions i.e. drop through limited height
\cite{wick}-\cite{ol2}. This has been made
possible by the advent of cheap and easily available electronic circuitry
capable of measuring accurately in milliseconds. In most of these
experiments, the body is released by either a lever or electromagnet. The
position of the release mechanism would be taken as s=0 and the initial velocity
on release would obviously be zero. A second sensor kept at a distance one
meter below the release point (s=0) would stop the timing circuit when the
body crosses it. The acceleration due to gravity is calculated using
eqn(\ref{eq1}). However, this scheme introduces an error in time 't' due to
the uncertainty in the time of release. In Lindemuth's \cite{jef} work, error was
introduced due to the body having an initial velocity by the time it arrived
to the first sensor that initiated the clock. In all schemes discussed,
there was an error either in the measurement of time of fall or knowledge of
initial velocity. Also, these experiments relay on the a prior knowledge/
validity of eqn(\ref{eq1}). To overcome this problem a method devised in use
is to allow a metallic ball to drop in between two parallel wires in which
high tension current flows in pulses. Sparks generated by the traveling
ball is recorded on a photographic strip. In another method two test masses to 
required to determine acceleration due to gravity \cite{wick}. These techniques 
either require high (unsafe) voltages or the contraptions
require a certain amount of skill in designing which would not be easily
available in standard under-graduate colleges of developing countries.

\par We have designed a simple setup to measure the acceleration due to
gravity. It was designed keeping in mind minimum skill should be required to
set it up and use it as also the validity of eqn(\ref{eq1}) would not be taken for
granted. In the following sections we elaborate our setup designed by using 
LEDs, photodiodes, op amps and microprocessor and comment on the results we
obtained. 

\section{Apparatus}

\begin{floatingfigure}[r]{2.75in}
\begin{center}
\epsfig{file=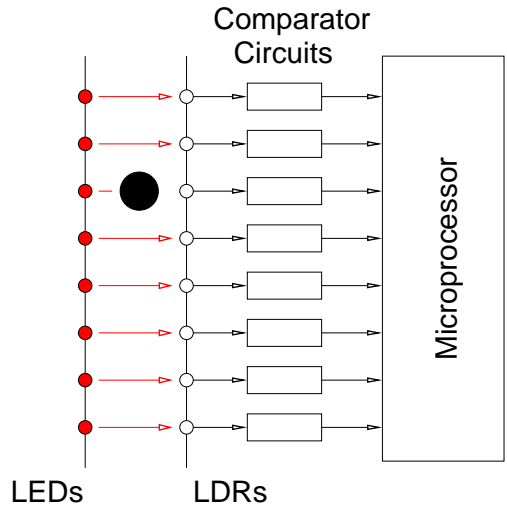, width=2.75in}
\caption{\sl Setup used to measure the time of fall of a falling object.} 
\label{fig:kalim1}
\end{center}
\vskip -0.4cm
\end{floatingfigure}

\par A hollow tube (which we shall hence refer to as the "{\it
drop tower}") of length 1meter housed eight LEDs (Light Emitting Diodes) and
eight LDRs( Light Dependent Resistances). They were arranged along the
length of the {\it drop tower} facing each other (see fig \ref{fig:kalim1}).
The distance between two neighboring LEDs (and in turn LDRs) was 10cm. The
first pair of LED-LDR was kept at a distance 10cm from the top edge of the
{\it drop tower}. This was to make sure that stray light from the room did
not lead to false triggering of the circuitry. 

\par
\begin{figure}[h]
\begin{center}
\epsfig{file=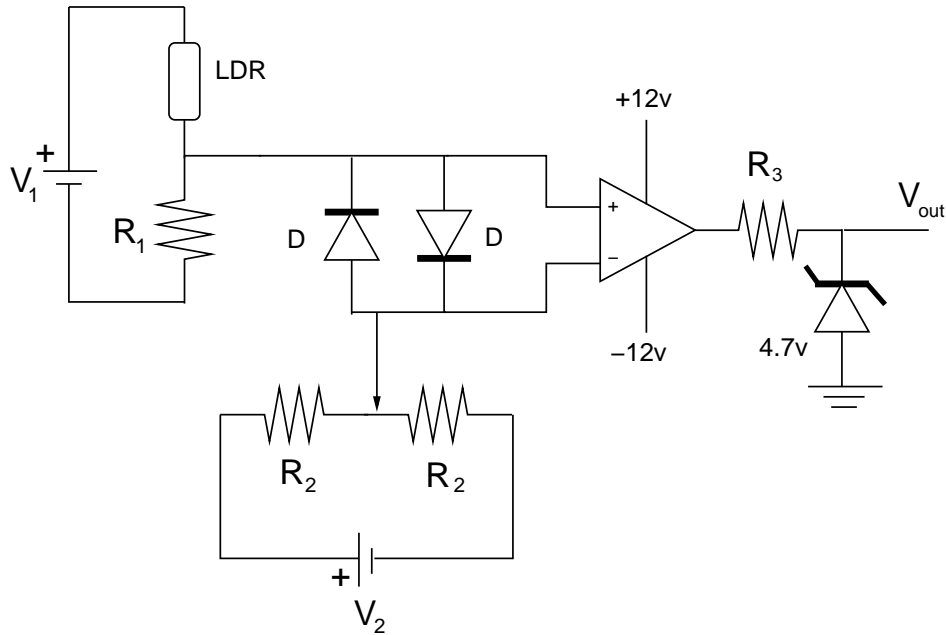,width=5in}
\caption{\sl Wave-shaping circuit used to convert variation of LDR's
resistance into square pulses. The DC supplies used were ${\rm V_1=5v}$ and
${\rm V_2=12v}$ and the resistances in the circuit were ${\rm
R_1=3.3K\Omega}$, ${\rm R_2=10K\Omega}$ and ${\rm R_3=1K\Omega}$. Diodes
used were the usual 1N4001 rectification diodes.} 
\label{fig:kalim2}
\end{center}
\vskip -0.4cm
\end{figure}

 As the name suggests, LDRs are light light sensitive devices and hence
are popularly used as optical sensors. Their resistance depends on the
amount of light falling on it. When light is incident on the LDR, it's
resistance is low. Thus, when a falling object cuts the lights path, no
light falls on the LDR and it's resistance increases. This variation in
resistance when converted into voltage can be (wave-)shaped into pulses which
then can be fed to a computer or microprocessor for identifying and storing
the information as to the instant the object passed the LDR. The required
circuitry used for wave-shaping is shown in fig(\ref{fig:kalim2}). The LDR
resistance is converted into a corresponding voltage using a voltage divider
circuit. The voltage across the resistance, ${\rm R_1}$, is high (${\rm
V_H}$) when light falls on the LDR since it's resistance diminishes.
Conversely, if the LDR is in shadow region it's resistance is high and the
voltage across ${\rm R_1}$ goes low (${\rm V_L}$). This voltage is given to
the op amp comparator circuit. The variable voltage is set at a value
between ${\rm V_L}$ and ${\rm V_H}$. This ensures that the output of the
LM741 op amp swings to +12v when light falls on the LDR and to -12v when
shadow falls on it. This voltage is made TTL/microprocessor compatible 
(vary between 0 and +5v) using a 4.7v zener diode.

\par Now we explain the final block of our setup which is a IC8085
microprocessor training kit. To follow in detail this section requires an
understanding of IC8085 and can be acquired from Gaonkar's book. A standard 
8085 microprocessor training kit comes with a programmable
input/output device (IC8155) which has as many as twenty two input pins
(divided into Port A with 8 pins, Port B with 8 pins and Port C with 6
pins). The output of each wave-shaping circuit can be fed to one of these
pins. We have only used eight pins of Port A. This was done keeping the
following in mind
\begin{itemize}
\item[(i)] Optimize the amount of hardware (wave-shaping circuits),
\item[(ii)] eight data points would be reasonably enough to fit a curve
reliability and
\item[(iii)] the circuit then could easily be extended for interfacing with
a computer since the computers parallel port can only input eight bits.
\end{itemize}

\begin{figure}[h]
\begin{center}
\epsfig{file=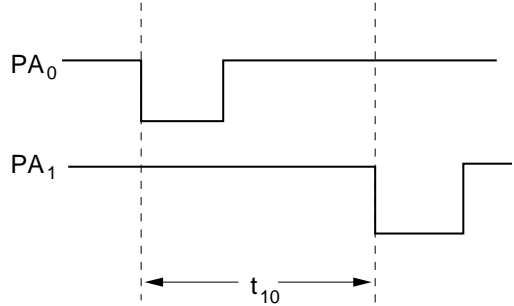, width=2.75in}
\caption{\sl Setup used to measure the time of fall of a falling object.} 
\label{fig:kalim3}
\end{center}
\vskip -0.4cm
\end{figure}

\par When no object cuts the light path, all the pins of Port A (of IC8155)
are high and informs the microprocessor by sending the word ${\rm FF_H}$
(the subscript 'H' means that it is the hexa representation of binary number
1111 1111). This word changes when a object cuts the light's path. For
example on cutting the first LED-light's path the word sent to the
microprocessor is 1111 1110 (${\rm FE_H}$) and when the object cuts the
second LED-light's path the microprocessor receives the word 1111 1101
(${\rm FD_H}$) and so forth as the object continuous to fall. The
microprocessor is programmed to run a counter (clock) from the negative edge
of one input signal to that of the second (see fig \ref{fig:kalim3}). On
receipt of the second input signal's negative edge, the microprocessor
stores the counter's count in it's memory and restarts the counter. This
contributes in an error (time measured would be lower than actual) in the 
measurement of the time of flight between the next two pair of LED-LDR. The
counts stored in the memory is converted to time in seconds using the
formula 
\begin{eqnarray}
t_1 &=& (30n+34)\times 0.326\mu s\nonumber\\
t_2\,\, to \,\, t_8 &=& (30n+41)\times 0.326\mu s\nonumber
\end{eqnarray}
Notice both the wave-shaping circuit and the software are designed to
minimize any delay in response by the cadmium sulphide LDR used. The
assembly language instruction set used for the project is listed at the end.

\section{Results and Discussions}

The {\it drop tower} was kept vertical and attempt was made to make sure
that the objects would fall in a straight line. A plumb line was used to
ascertain the alignment. For conducting this experiment we used four
different balls whose physical features, dimensions and mass are listed in
Table 1. 

\begin{table}[b]
\begin{center}
{\bf Table 1.}
\end{center}
\begin{center}
\begin{tabular}{||c|c|c|c||}\hline\hline
Ball Name & Description & Mass (grams) & Radius (cm) \\
\hline
Ball 1 & Smooth Rubber Ball &	19.00		& 1.4\\
Ball 2 & Ping Pong Ball &	1.75		& 1.9\\
Ball 3 & Nylon Ball &	5.50		& 1.9\\
\hline\hline
\end{tabular} 
\end{center}
\end{table}

The balls were dropped from the top edge of the {\it drop tower}.
We made sure that the ball in use would cut all eight light beams. As soon
as the ball crosses the first LED, the microprocessor starts it's clock.
This point would act as the co-ordinates (0,0). The microprocessor stores
the counter's count as the ball passes the second LED (the count in turn
would be used to calculate time). This gives the co-ordinates 
(${\rm t_{21}}$, 0.1) 
where the subscript '21' indicates this as the time taken for
the fall between the first and second LED. Since, the LEDs are 10cm apart,
in meters we have s=0.1cm. In this manner, eight data points are generated
and stored in the microprocessor. The experiment was repeated as many as 200
times with each ball. 
\begin{figure}[h]
\begin{center}
\epsfig{file=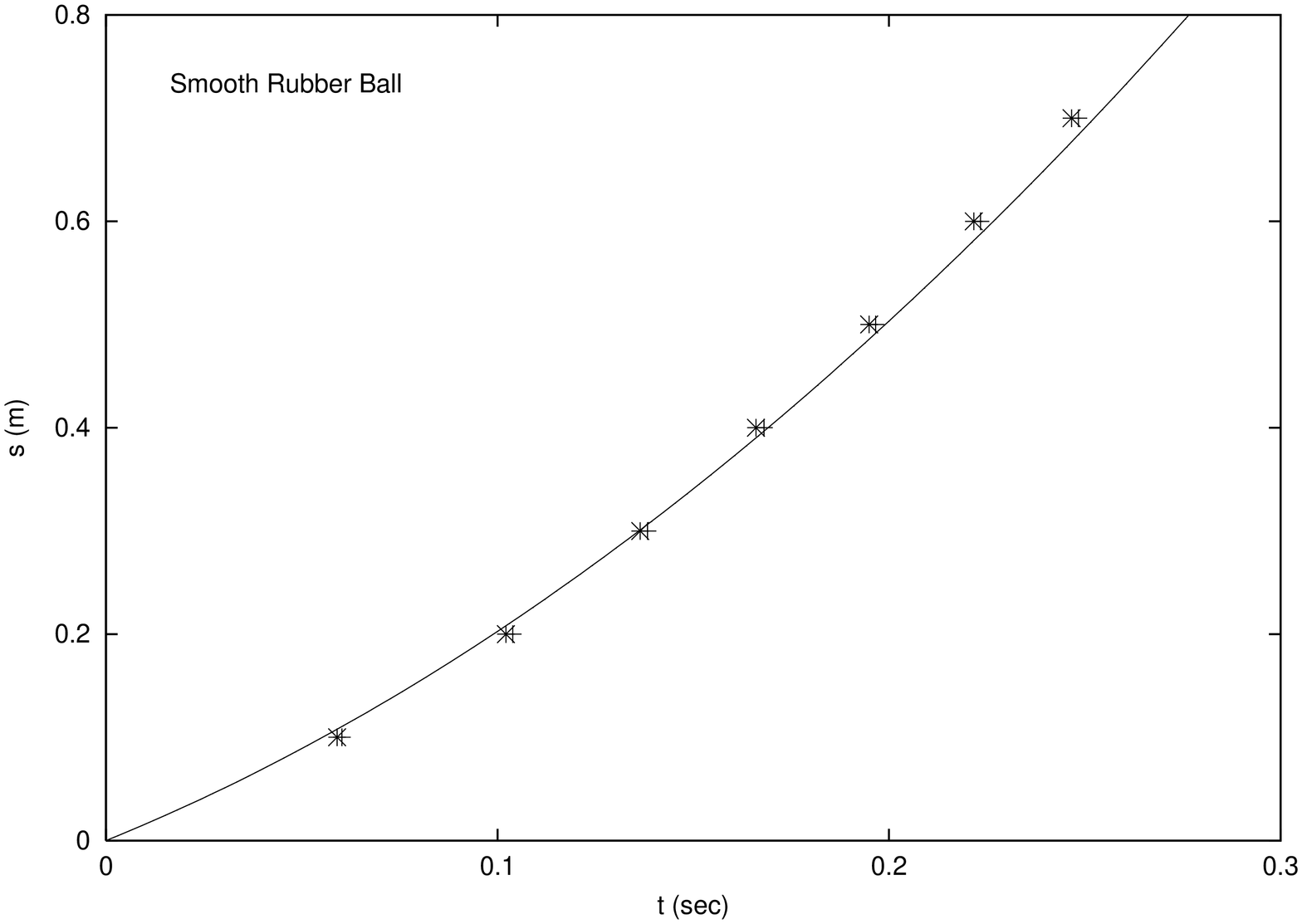, width=1.95in,angle=-90}
\hfil
\epsfig{file=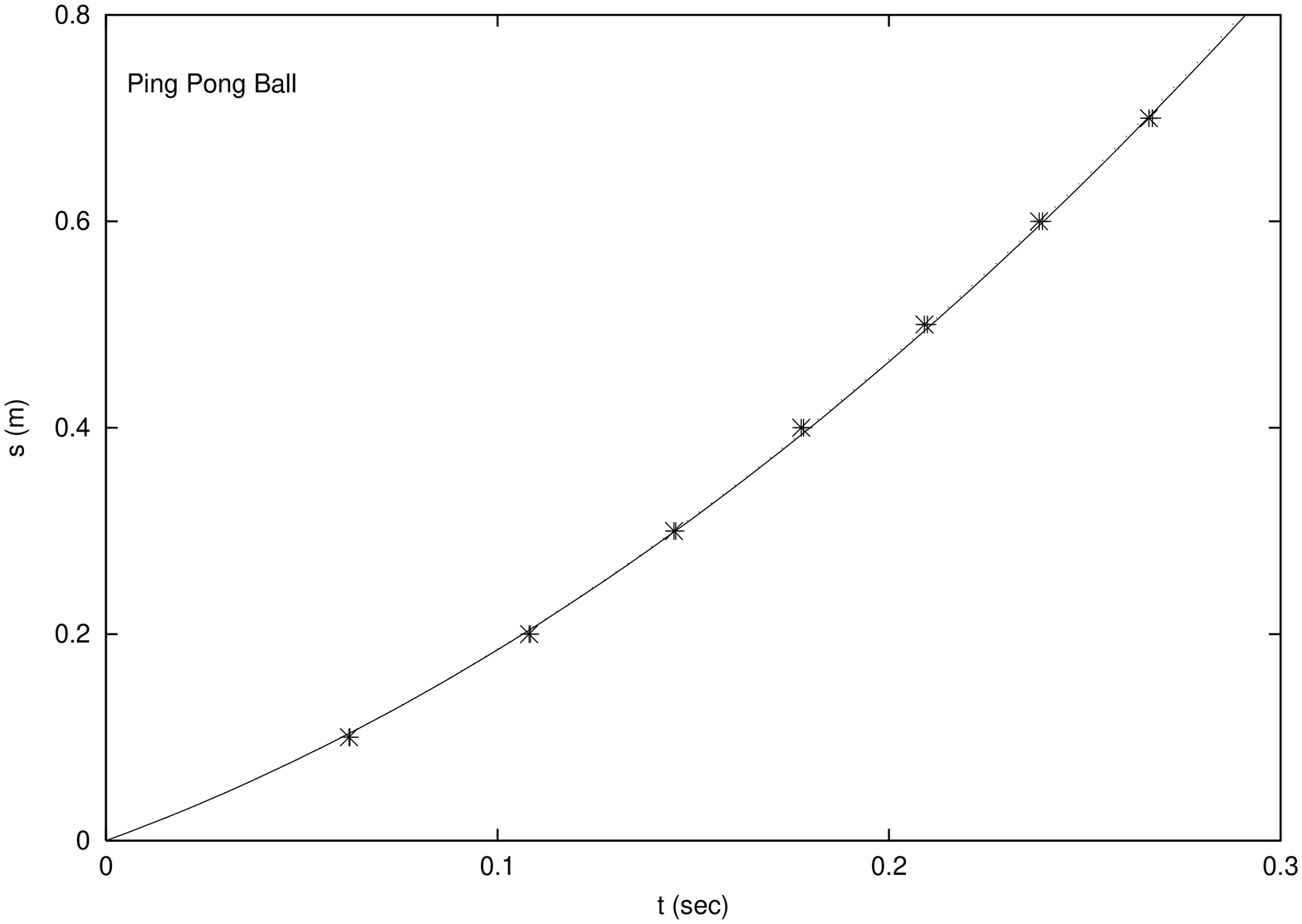, width=1.95in, angle=-90}
\caption{\sl The time versus position graph for Ball 1 (smooth rubber ball) and 
Ball 2 (ping-pong ball).} 
\label{fig:kalim4}
\end{center}
\vskip -0.4cm
\end{figure}

\par Fig(\ref{fig:kalim4}) shows two curves each for the results obtained
with Ball 1 and Ball 2. As can be seen the data lie on a parabola. However,
before fitting a second polynomial that would represent eq(\ref{eq1}), one
has to remember that eq(\ref{eq1}) represents a object falling freely under
the action of gravity. No other forces should be acting on it. However,
there are two possible forces acting in the opposite direction to gravity
when the object is dropped, namely forces of buoyancy and air resistance. The
contribution due to air resistance is proportional to the square of the
velocity with which the object is moving. The general
equation of motion in this case is given as
\begin{eqnarray}
m{d^2s \over dt^2}=(m-m_{air})g-kv^2\nonumber
\end{eqnarray}
where ${\rm m_{air}}$ is the mass of the air displaced by the ball and 'k'
is the proportionality constant. Since, the dimensions of the balls
are small, the volume and in turn the mass of air displaced is very small.
Hence, contributions of the buoyancy force can be neglected. Hence, the
equation of motion reduces to
\begin{eqnarray}
m{d^2s \over dt^2}=mg-kv^2\label{eqn6}
\end{eqnarray}
Before proceeding, for completeness it should be stated here that the force 
acting as a resistance on the falling body can either be proportional to
it's velocity (given by Navier-Strokes equation, ${\rm F=6\pi \mu rv}$) or the 
square of the velocity. This depends if motion of the viscous medium around
the falling body remains laminar or not and is denoted by the Reynolds
number (R)
\begin{eqnarray}
R={d\rho v \over \mu}\nonumber
\end{eqnarray}
where this number characterises the media (viscosity, ${\rm \mu}$ and
density, ${\rm \rho}$) through which the body (of diameter, d) falls with
velocity, v. In a very viscous media (${\rm \mu}$ large), the body would
face large resistance and would fall slowly (v small). Thus the Reynolds
number would be less than one. Under such conditions the NAvier-Strokes
equation should be considered. However, as in our case, the body is falling
through (dry) air, whose viscosity is small (around ${\rm 1.9\times
10^{-5}kg/m.sec}$) with density ${\rm 1.225kg/m^3}$. The Reynolds number
works out be ${\rm \approx 3000}$, i.e. ${\rm R \gg 1}$. The drag force
acting on the body is then given as
\begin{eqnarray}
F &=& {1\over 2}\rho A C_Dv^2\nonumber\\
&=& kv^2\nonumber
\end{eqnarray}
or ${\rm k={1 \over 2}C_D\rho A}$, where 'A' is the area of cross-section
and ${\rm C_D}$ is the drag coefficient. The drag coefficient depends on the
shape, surface characteristics and the Reynolds number \cite{weiss}. The value 
of ${\rm C_D}$ for smooth spheres as a function of Reynolds number has been 
listed in Landau and Lifshitz book \cite{book}. For R=3000, ${\rm C_D \approx 
0.5}$. For the radii of the balls used in our experiment (see Table 1), k works 
out to be ${\rm \approx 3.8\times 10^{-4}(Ns^2/m^2)}$. However, we shall not use 
this value immediately but at the end compare our experimentally determined 
values with this theoretical value. 

\par The solution of eqn(\ref{eqn6}) and it's power approximation has been
detailed by Lindemuth and is given as
\begin{eqnarray}
s &=& \left({v_T^2 \over g}\right)ln\left[cosh{gt \over v_T}+{u \over
v_T}sinh{gt \over v_T}\right]\nonumber\\
&\approx & ut+{1 \over 2}g\left(1-{u^2 \over v_T^2}\right)t^2
-{g^2u \over 3v_T}\left(1-{u^2 \over v_T^2}\right)t^3\label{eqn3}
\end{eqnarray}
where 
\begin{eqnarray}
v_T=\sqrt{mg \over k}\nonumber
\end{eqnarray}
is the terminal velocity the falling body would achieve when it moves with
no acceleration, i.e. net force acting on it is zero in the case of free
fall else when all other higher terms cancel out. The coefficient of 
${\rm t^3}$ represents the rate of change in acceleration (dg/dt). 

Thus, under the conditions in which this experiment was done, the data
points were fit with a third order polynomial (eqn \ref{eqn3}). Curves
were fit to these data points using standard and freely available software
"Curxpt v3.0". 

\begin{table}[h]
\begin{center}
{\bf Table 2.}
\end{center}
\begin{center}
\begin{tabular}{||c|c|c|c|c|c||}\hline\hline
Ball Name & & ${\rm u_1}$ (m/s) & g (${\rm m/s^2}$)& dg/dt (${\rm m/s^3}$) & co-relation \\
\hline
Ball 1 & I & 1.559 & 9.7998 & 8.0e-9 & 0.998\\ \cline{2-6}
 & II & 1.559 & 9.7998 & 4e-9 & 0.998\\ \cline{1-6}
Ball 2 & I & 1.363 & 9.798 & 0.643 & 0.999\\ \cline{2-6}
 & II & 1.368 & 9.7745 & 0.731 & 0.999\\\cline{1-6}
Ball 3 & I & 1.549 & 9.7998 & 0.418 & 0.998\\ \cline{2-6}
 & II & 1.445 & 9.7822 & 0.268 & 0.999\\ \cline{1-6}
\hline\hline
\end{tabular} 
\end{center}
\end{table}

All the curves for the 200 odd experiments done with each
ball lie within the two curves shown in fig(\ref{fig:kalim4}). The
coefficient of t, ${\rm t^2}$ and ${\rm t^3}$ as returned by the software was 
used to determine the ball's velocity (${\rm u_1}$) as it crosses the first 
sensor where the ball has traveled through 10+cm (10cm from the drop tower's 
top edge to the first sensor and the unknown height from the edge from where 
the ball was dropped) and it's acceleration (g). Results for the graphs shown 
in fig(\ref{fig:kalim4}) are listed in Table 2. The error in 'g' is 1 in part 
of ${\rm 10^3}$ with as small as ${\rm +7\mu sec}$ (max) error in the 
measurement of time of flight. The acceleration due to gravity in Delhi can 
be calculated using the equation 
\begin{eqnarray}
g=9.780327[1+0.0053024sin^2L-0.000058sin^22L]\label{eq2}
\end{eqnarray}
where 0.499455 (${\rm 28^o 37'}$N) is the latitude (L) of Delhi in radian.
This gives the acceleration due to gravity in Delhi as ${\rm 9.7918m/s^2}$.
Our experimental value of 'g' is in good agreement with the theoretical
value. Notice that the velocity (${\rm u_1}$) is different in each case 
(see Table 2). This is expected since even with the best of care taken, 
the balls would have been dropped from different heights (position from the 
{\it drop tower's} edge).

\par The power series of eqn(\ref{eqn3}) suggests that a knowledge of the
coefficients should also enable us to estimate the objects terminal
velocity. However, the power approximation leads to the situation that if
the body's initial velocity is zero then the terms associated with terminal
velocity also collapses. Also, curve fitting the original solution relating
distance to time proves tedious due to the existence of logaritm and cosine
hyperbolic terms. However, the functional relation between velocity and time
for a body falling through a viscous medium is given as \cite{jef}
\begin{eqnarray}
v=v_Ttanh\left[{g \over v_T}t+tanh^{-1}\left({u \over v_T}\right)\right]\nonumber
\end{eqnarray}
When the body's initial velcoity is zero, the above expression reduces to
\begin{eqnarray}
v=v_Ttanh\left[{g \over v_T}t\right]\label{eqn5}
\end{eqnarray}
One can either use the coefficients 'g' and 'dg/dt' in
\begin{eqnarray}
v=gt-3{dg \over dt}t^2\nonumber
\end{eqnarray}
to generate data for curve fitting eqn(\ref{eqn5}) or find the gradient
along the curves shown in fig(\ref{fig:kalim4}). Both methods would introduce
an error abid small since the fitting as represented by co-relation factor
(see Table 2) was very good. The terminal velcoities thus evaluated is
listed in Table 3. Before concluding, it would be interesting to compute the
time a body would take to attain terminal velcoity and the distance through
which it would fall to attain this velocity. For this one would have to
solve the quadratic equation
\begin{eqnarray}
v_T=gt-3{dg \over dt}t^2\nonumber
\end{eqnarray}
and substitute the time taken thus obtained in the equation
\begin{eqnarray}
s={1 \over 2}gt^2-{dg \over dt}t^3\nonumber
\end{eqnarray}

\begin{table}[t]
\begin{center}
{\bf Table 3.}
\end{center}
\begin{center}
\begin{tabular}{||c|c|c|c|c|c||}\hline\hline
Ball Name & & ${\rm v_T}$ (m/s) & t (sec) & s (m) & ${\rm k=mg/v_T^2}$
 ${\rm \times 10^{-4} (Ns^2/m^2)}$\\
\hline
Ball 1 & I & 35.11 & 3.58 & 62.9 & 1.27\\ \cline{2-6}
 & II & 35.11 & 3.58 & 62.9 & 1.27\\ \cline{1-6}
Ball 2 & I & 6.32 & 0.58 & 1.77 & 4.3\\ \cline{2-6}
 & II & 5.89 & 0.53 & 1.52 & 4.92\\\cline{1-6}
Ball 3 & I & 7.97 & 0.74 & 2.87 & 8.47\\ \cline{2-6}
 & II & 9.97 & 0.94 & 4.6 & 5.4\\ \cline{1-6}
\hline\hline
\end{tabular} 
\end{center}
\end{table}

\par The data of Table 3 indicates that the terminal velocity, the time in
which the object attains this velocity and the distance through which it has
to travel for attaining the same all strongly depends on the objects mass.
The drag coefficient also depends on the surface area (and hence
diameter) hence comparison was sort to be done on balls with similar
diameter and different mass. Obtaining balls of different mass and similar
dimensions was not easy. However, the designed experimental setup proved handy 
in estimating the acceleration due to gravity. Above which, it also proved to 
be a rich source of information on the motion of an object while falling. 
Providing ample scope of and experience in data analysis, the setup gives an 
idea of the terminal velocity, time taken to achieve it and the distance through 
which the body has to travel before attaining it. All this is possible eventhough 
the length of the drop tower is less than the distance through which object has
to fall to attain ${\rm v_T}$. On a final note it must be mentioned that this 
project was done in a graduate college laboratory, with a small budget of 
${\rm \$20}$.

\section*{Acknowledgements}

The authors would like to express their gratitude to the lab technicians of
the Department of Physics and Electronics, SGTB Khalsa College, for the help
rendered in carrying out the experiment.

\vfil \eject
\pagebreak
\begin{table}[t]
\begin{center}
{\sl Program required.}
\end{center}
\begin{center}
\begin{tabular}{||c|c|c||}\hline\hline
Mem. add. & Hex Code & Instruction \\
\hline
C100 &	31 &	LXI SP		\\
C101 &	00 &	00		\\
C102 &	C6 &	C6		\\
C103 &	21 &	LXI H		\\
C104 &	00 &	00		\\
C105 &	C0 &	C0		\\
C106 &	3E &	MVI A		\\
C107 &	00 &	00		\\
C108 &	D3 &	OUT 08 (CR)	\\
C109 &	08 &	08		\\
C10A &	DB &	IN 09 (PA)	\\
C10B &	09 &	09		\\
C10C &	A6 &	ANA M		\\
C10D &	C2 &	C2		\\
C10E &	0A &	0A		\\
C10F &	C1 &	C1		\\
C110 &	23 &	INX H		\\
C111 &	11 &	LXI D (initiate counter)\\
C112 &	00 &	00		\\
C113 &	00 &	00		\\
C114 &	13 &	INX D		\\
C115 &	DB &	IN 09 (PA)	\\
C116 &	09 &	09		\\
C117 &	A6 &	ANA M		\\
C118 &	C2 &	JNZ		\\
C119 &	11 &	11		\\
C11A &	C1 &	C1		\\
C11B &	23 &	PUSH D		\\
C11C &	D5 &	INX H		\\
C11D &	C3 &	JMP		\\
C11E &	11 &	11		\\
C11F &	C1 &	C1 		\\
\hline\hline
\end{tabular} 
\end{center}
\end{table}


\begin{table}[b]
\begin{center}
{\sl Data used by main program.}
\end{center}
\begin{center}
\begin{tabular}{||c|c||}\hline\hline
Mem. add. & Data\\
\hline
C000 &	01 \\
C001 &	02 \\
C002 &	04 \\
C003 &	08 \\
C004 &	10 \\
C005 &	20 \\
C006 &	30 \\
C007 &	80 \\
\hline\hline
\end{tabular} 
\end{center}
\end{table}

\end{document}